\documentclass[twocolumn,showpacs,pre,floatfix,amsmath,amssymb,notitlepage,superscriptaddress]{revtex4-1}
\usepackage{epsfig}
\usepackage{subfigure}
\usepackage{amsmath}
\usepackage{color}
\usepackage{amssymb}
\usepackage{setspace}
\usepackage{graphicx}
\usepackage{dcolumn}
\usepackage{bm}
\usepackage{times}
\usepackage{enumerate}
\usepackage{footnote}
\input epsf

\usepackage{algorithm}	
\usepackage{algpseudocode}

\newcommand{\drt}[1]{{\color{blue}#1}}

\newcommand{\appropto}{\mathrel{\vcenter{
  \offinterlineskip\halign{\hfil$##$\cr
    \propto\cr\noalign{\kern2pt}\sim\cr\noalign{\kern-2pt}}}}}

\graphicspath{{figures/}}

\begin{document}

\author{Pranick R. Chamlagai}
\affiliation{Department of Mathematics, Trinity College, Hartford, CT 06106, USA}

\author{Dane Taylor}
\affiliation{Department of Mathematics, University at Buffalo, State University of New York, Buffalo, NY 14260, USA}

\author{Per Sebastian Skardal}
\email{persebastian.skardal@trincoll.edu} 
\affiliation{Department of Mathematics, Trinity College, Hartford, CT 06106, USA}

\title{Grass-roots optimization of coupled oscillator networks}

\begin{abstract}
Despite the prevalence of biological and physical systems for which synchronization is critical, existing  theory for optimizing synchrony depends on global information and does not sufficiently explore local mechanisms that enhance synchronization. Thus, there is a lack of understanding for the self-organized, collective processes that aim to optimize/repair synchronous systems, e.g., the dynamics of paracrine signaling within cardiac cells. Here we present ``grass-roots'' optimization of synchronization, which is a multiscale mechanism in which local optimizations of smaller subsystems cooperate to collectively optimize an entire system. Considering models of cardiac tissue and a power grid, we show that grass-roots-optimized systems are comparable to globally optimized systems, but they also have the added benefit of being robust to targeted attacks or subsystem islanding. Our findings motivate and support further investigation into the physics of local mechanisms that can support self-optimization for complex systems.
\end{abstract}


\maketitle

\section{Introduction}\label{sec:01}

The ability for large systems of dynamical units to self-organize and produce robust collective behavior continues to drive a large body of research~\cite{Pikovsky2003,Strogatz2004}. Applications include cardiac dynamics~\cite{Bychkov2020JACC},  brain dynamics~\cite{Kopell2000PNAS}, cell signaling~\cite{Prindle2012Nature}, and power grids~\cite{Rohden2012PRL}. Weak synchronization and desynchronization events often lead to pathological behavior, e.g., spiral wave breakup in cardiac tissue~\cite{Fenton2002Chaos,Panfilov2007PNAS} and black outs in power grids~\cite{Dorfler2013PNAS}, thereby motivating {\it optimized} systems for strong, robust synchronization. 

While man-made systems such as power grids can be designed using global structural and dynamical information~\cite{Pecora1998PRL,Nishikawa2006PRE}, such information is likely unavailable to biological processes that are known to rely on local interactions, such as cell-to-cell paracrine signaling among cardiac cells \cite{hodgkinson2016emerging}. 
While a great deal is known about how biological systems function, comparatively little is understood about the self-optimization processes responsible for constructing and maintaining/repairing such systems. Moreover, it is reasonable to hypothesize that optimization is itself a collective, coordinated behavior. A stronger theoretical understanding of mechanisms for collective self-optimization may deepen our understanding of diverse biological systems and has the potential to revolutionize the way we engineer systems--or rather, design systems to engineer themselves. Collective optimizations constitute an under-explored family of collective behavior, and there is a lack of multiscale optimization theory to provide insight into how local optimizations might coordinate to globally optimize both synchronous and other kinds of systems.

In this paper, we explore {\it grass-roots optimization} for coupled oscillator networks, whereby the parallel optimization of smaller subsystems can be coordinated to collectively optimize the global synchronization properties of the entire system. In general, {\it subsystems} can be defined in a variety of ways:  community structure~\cite{Girvan2002PNAS}, spatially distinct regions in a geometric network~\cite{Barthelemy2011PhysRep},  or other partitions of a network after a geometric embedding~\cite{Coiffman2005PNAS}. Such locally defined subsystems are consistent with the tissue microenvironments that emerge via paracrine signaling in cardiac tissue undergoing stem cell therapy~\cite{hodgkinson2016emerging}. Our main finding is an intuitive multiscale mechanism for grass-roots optimization of synchronization that involves two steps: {\it local subsystem optimization}, whereby subsystems are independently optimized in parallel; and {\it global subsystem balancing}, whereby the subsystems are balanced with one another. Grass-roots optimization coordinates two seemingly contrasting ideas whereby (i) optimized networks tend to connect dissimilar oscillators\drt{,} and (ii) similarity between two oscillators promotes their entrainment. Specifically, a multiscale approach allows subsystems to be treated as near-identical ``macro-oscillators'' while preserving and taking advantage of heterogeneity on a microscopic scale. 

We demonstrate the utility of grass-roots optimization across a range of networks where subsystems arise naturally: random networks with communities, a power grid, and a geometric network model. Moving beyond phase oscillators, we also use a nonlinear cardiac pacemaker model for which we optimize voltage and gating variables of pacemaker cells~\cite{Djabella2007IEEE}. In addition to successfully optimizing synchronization dynamics, grass-roots-optimized systems also have the added benefit of being more robust to subsystem dismantling under a targeted attack or intentional islanding than globally optimized systems. These experiments highlight grass-roots optimization as a viable mechanism by which diverse types of systems can robustly self-optimize, providing a plausible mechanism to support biological systems as well as decentralized engineering strategies for complex man-made systems.

\begin{figure*}[t]
\centering
\epsfig{file =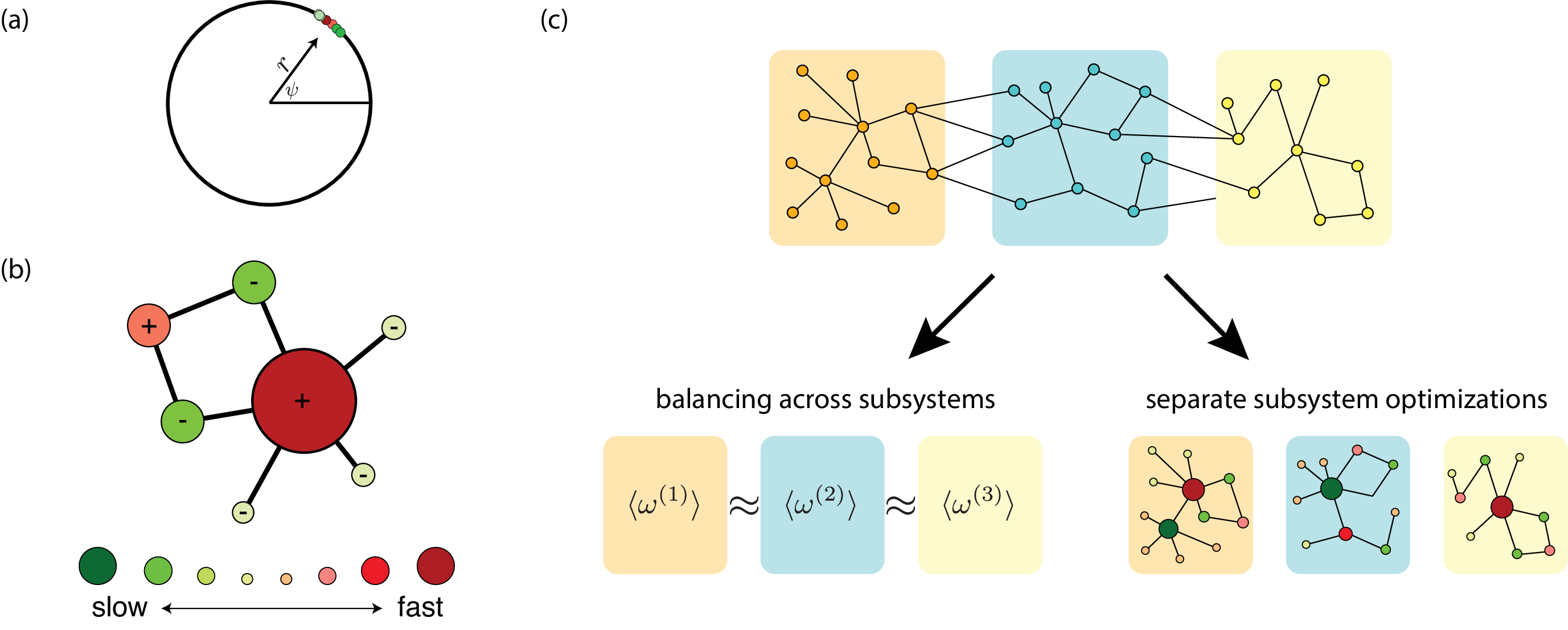, clip =,width=0.90\linewidth }
\caption{{\it Grass-roots optimization for a network of heterogenous phase oscillators}. (a) Visualization of Kuramoto order parameter $r\in[0,1]$ and mean field $\psi\in[0,,2\pi)$ for a set $\{\theta_j\}$ of oscillator phases with $\theta_j\in[0,2\pi)$. Strong phase synchronization occurs when $\theta_i\approx\theta_j$ for any $i$ and $j$, which yields $r\approx 1$. (b) Synchrony-optimized networks that maximize $r$ can be obtained using the synchrony alignment function (SAF)~\cite{Skardal2014PRL}, which reveals two microscale, intuitive mechanisms that promote synchronization: positive correlations between an oscillator's natural frequency $\omega_i$ and its associated node degree $d_i$; and negative correlations among the frequencies $\omega_i$ and $\omega_j$ of neighboring oscillators $i$ and $j$. (c) Here, we develop grass-roots optimization to reveal a multiscale mechanism for optimization with two steps: the mean frequency $\langle \omega^{(s)}\rangle$ within each subsystem $s$ is balanced across the subsystems; and    subsystems are separately optimized.}\label{fig1}
\end{figure*}

The remainder of this paper is organized as follows. In Sec.~\ref{sec:02} we summarize some preliminaries and present our main result: a grass-roots optimization framework for network synchronization. In Sec.~\ref{sec:03} we present numerical experiments to highlight the effectiveness of this framework, and in Sec.~\ref{sec:07} we conclude with a discussion of our results. 

\section{Main Results}\label{sec:02}

Here, we review a synchronization optimization framework in Sec.~\ref{sec:2a}, present a local approximation theory for optimization in Sec.~\ref{sec:2b}, and develop a grass-roots optimization framework for synchronization in Sec.~\ref{sec:2c}. We visualize visualize synchrony optimization and our grass-roots approach in Fig.~\ref{fig1}.

\subsection{The Synchrony Alignment Function (SAF)}\label{sec:2a}

We begin by reviewing the synchrony alignment function for the optimization of networks of heterogeneous oscillators. Consider a network of coupled, heterogeneous phase oscillators whose dynamics are given by
\begin{align}
\dot{\theta}_i=\omega_i+K\sum_{j=1}^NA_{ij}H(\theta_j-\theta_i),\label{eq:01}
\end{align}
where $\theta_i$ and $\omega_i$ are the phase and natural frequency of oscillator $i=1,\dots,N$, parameter $K$ is the global coupling strength, network structure is encoded in an adjacency matrix $A$, and $H$ is a $2\pi$-periodic coupling function. Here, we focus on the case of unweighted, undirected networks with $A_{ij} = 1$ if oscillators $i$ and $j$ are connected and $0$ otherwise, although these properties may be relaxed without much trouble. We also use classical Kuramoto coupling~\cite{Kuramoto}, i.e., $H(\cdot)=\sin(\cdot)$, but emphasize that one may choose other functions $H$ provided that $H'(0)>0$ and $H(\Delta\theta)=0$ for some $\Delta\theta$ near zero. Notably, phase oscillator models such as Eq.~(\ref{eq:01}) have been found to be suitable models for naturally-occuring phenomena such as chromosomal coordination~\cite{Rajapakse2009PNAS} and integrate and fire dynamics of cardiac pacemakers~\cite{Politi2015PRE}, as well as mechanical systems such as power grids~\cite{Porco2013,Skardal2015SciAdv}.

 The degree of synchronization is measured by the magnitude $r\in[0,1]$ of the Kuramoto order parameter 
 \begin{align}
 re^{i\psi}=N^{-1}\sum_{j=1}^Ne^{i\theta_j},
 \end{align} 
 which we illustrate for a strongly synchronized state in Fig.~\ref{fig1}(a). 
By linearizing around the synchronized state one obtains
\begin{align}
r\approx1-\frac{J(\bm{\omega},L)}{2K^2},\label{eq:022}
\end{align}
where
\begin{align}
J(\bm{\omega},L)=\frac{1}{N}\sum_{j=2}^N\frac{\langle\bm{v}^j,\bm{\omega}\rangle^2}{\lambda_j^2}\label{eq:02}
\end{align}
is the Synchrony Alignment Function (SAF)~\cite{Skardal2014PRL}. The SAF utilizes the alignment of the natural frequencies $\bm{\omega}$ with the eigenvalues $\{\lambda^j\}_{j=1}^N$ and eigenvectors $\{\bm{v}^j\}_{j=1}^N$ of the combinatorial Laplacian, $L=D-A$, where $D=\text{diag}(k_1,\dots,k_N)$ is a diagonal matrix that encodes the nodal degrees, $k_i=\sum_{j=1}^NA_{ij}$. Synchronization is optimized (i.e., $r$ is maximized) by minimizing $J(\bm{\omega},L)$, which may be done by aligning $\bm{\omega}$ with the eigenvectors of $L$ that are associated with larger eigenvalues. 
The SAF framework has been utilized across several optimization scenarios, including undirected and directed networks~\cite{Skardal2014PRL,Skardal2016Chaos}, finding optimal perturbations and network rewirings~\cite{Taylor2016SIAM,Arola2021Chaos}, synchronizing phase-coherent chaotic oscillator networks~\cite{Skardal2017Chaos}, and dealing with frequency uncertainty~\cite{Skardal2019SIAM}.

Minimizing the SAF with $\omega\propto v^N$ also reveals intuitive key properties of synchrony optimized systems including degree-frequency correlations and anti-correlations between neighboring frequencies~\cite{Skardal2014PRL}. These are illustrated in Fig.~\ref{fig1}(b), as neighboring oscillators tend to have frequencies with opposite signs, and high-degree nodes tend to be substantially faster or slower (i.e., larger or smaller natural frequencies) with respect to the average. While such local properties are associated with optimization, they alone do not guarantee it, nor do they offer insight toward mesoscale/multiscale properties and mechanisms enabling collective optimization.

\subsection{Local Approximation  for Networks with Two Subsystems}\label{sec:2b}

Here, we present a local approximation of the SAF, which which we will use to identify a multiscale mechanism underlying grass-roots optimization. For simplicity,  here we only consider the case of a network with two subsystems, leaving further generalization to the Appendix. The three subsystem case is detailed in Appendix~\ref{app:01}, and generalization to an arbitrary number of subsystems is discussed in Appendix~\ref{app:02}.

Writing the adjacency matrix as $A=\begin{bmatrix}A^{(1)} & B^{(12)} \\ B^{(12)T} & A^{(2)}\end{bmatrix}$, where $A^{(1)}\in\mathbb{R}^{N_1\times N_1}$, $A^{(2)}\in\mathbb{R}^{N_2\times N_2}$, $B^{(12)}\in\mathbb{R}^{N_1\times N_2}$, and $N_1$ and $N_2$ are the sizes of the respective subsystems, the Laplacian is given by $L=L_0+L_B$, where $L_0=\text{diag}(L^{(1)}, L^{(2)})$, $L^{(1,2)}=D^{(1,2)}-A^{(1,2)}$, and $L_B=\begin{bmatrix}D_{B^{(12)}} & -B^{(12)} \\ -B^{(12)T} & D_{B^{(12)T}}\end{bmatrix}$ with diagonal matrices $D_{B^{(12)}}$ and $D_{B^{(12)T}}$ whose entries are row sums of $B^{(12)}$ and $B^{(12)T}$, respectively. We assume $B^{(12)}$ to be sparser than $A^{(1)}$ and $A^{(2)}$ so that $\left\|L_B\right\|\ll\left\| L_0 \right\|$ under a suitable matrix norm (e.g., the Frobenius norm). We then define $\Delta L=(\|L_0\|/\|L_B\|)L_B$ so that $L(\epsilon)=L_0+\epsilon\Delta L$ recovers the original network structure for the choice $\epsilon=\|L_B\|/\|L_0\|\ll1$.

Next, we discuss the spectral properties of $L_0$. Since this matrix encodes the two subsystems in isolation, its eigenvalue spectrum is the union of the eigenvalue spectrum of $L^{(1)}$ and $L^{(2)}$. Specifically, ordering the eigenvalues of $L^{(1)}$ and $L^{(2)}$, respectively, $0=\mu_1<\mu_2\le\cdots\le\mu_{N_1}$ and $0=\nu_1<\nu_2\le\cdots\le\nu_{N_2}$ (where we assume that the subsystems are themselves connected), this implies that $L_0$ has two zero eigenvalues, $\lambda_1=\lambda_2=0$, with the rest positive, so that the nullspace of $L_0$ requires some care. Rather than choosing eigenvectors $\bm{v}^1\propto[\bm{1},\bm{0}]^T$ and $\bm{v}^2\propto[\bm{0},\bm{1}]^T$, whose entries are constant within one subsystem and zero within the other, it is advantageous to instead choose $\bm{v}^1=\frac{1}{\sqrt{N}}[\bm{1},\bm{1}]^T$ and $\bm{v}^2=\frac{\sqrt{N_1N_2}}{N}[\bm{1}/N_1,-\bm{1}/N_2]^T$ so that $\bm{v}^1$ is independent of $\epsilon$ and characterizes the nullspace of $L(\epsilon)$, and $\bm{v}^2$ is associated with an eigenvalue that converges to 0 as $\epsilon\to0$ but is strictly positive for $\epsilon>0$. The other $N-2$ eigenvectors of $L_0$ are given by $\{\bm{v}^j\}_{j=3}^N=\left\{[\bm{u}^j,\bm{0}]^T\right\}_{j=2}^{N_1}\bigcup\left\{[\bm{0},\bm{x}^j]^T\right\}_{j=2}^{N_2}$, where $\{\bm{u}^j\}_{j=1}^{N_1}$ and $\{\bm{x}^j\}_{j=1}^{N_2}$ are the eigenvectors of $L^{(1)}$ and $L^{(2)}$.

Considering $0<\epsilon\ll1$, each eigenvalue of $L(\epsilon)$ varies continuously with $\epsilon$~\cite{Kato2013}, so we may write $\lambda_j(\epsilon)=\lambda_j+\epsilon\delta\lambda_j^{(1)}+\epsilon^2\delta\lambda_j^{(2)}+\mathcal{O}(\epsilon^3)$. We similarly assume $\bm{v}^j(\epsilon)=\bm{v}^j+\epsilon\delta\bm{v}^{j(1)}+\epsilon^2\delta\bm{v}^{j(2)}+\mathcal{O}(\epsilon^3)$. Since $\lambda_2(\epsilon)\ll1$ and $\lambda_j(\epsilon)\sim1$ for $j=3,\dots,N$, the term associated with $j=2$ needs to be treated separately, so we write
\begin{align}
J(\bm{\omega},L(\epsilon))=\frac{1}{N}\frac{\langle\bm{\omega},\bm{v}^2(\epsilon)\rangle^2}{\lambda_2^2(\epsilon)}+\frac{1}{N}\sum_{j=3}^N\frac{\langle\bm{\omega},\bm{v}^j(\epsilon)\rangle^2}{\lambda_j^2(\epsilon)}.\label{eq:04}
\end{align}
Upon expanding the $N-1$ terms contributing to the SAF in Eq.~(\ref{eq:04}), we find that they all take a similar form except for a factor of $\epsilon$,
\begin{widetext}
\begin{align}
&\left(\frac{\langle\bm{\omega},\bm{v}^j(\epsilon)\rangle}{\lambda_j(\epsilon)}\right)^2=\epsilon^{\alpha_j}\left(\frac{\langle\bm{\omega},\bm{v}^j\rangle^2}{(\lambda_j)^2}\right)+\epsilon^{1+\alpha_j}\left(\frac{2\langle\bm{\omega},\bm{v}^j\rangle\langle\bm{\omega},\delta\bm{v}^{j(1)}\rangle}{(\lambda_j)^2}-\frac{2\delta\lambda_j^{(1)}\langle\bm{\omega},\bm{v}^j\rangle^2}{(\lambda_j)^3}\right)\nonumber\\
&+\epsilon^{2+\alpha_j}\left(\frac{\langle\bm{\omega},\delta\bm{v}^{j(1)}\rangle^2+2\langle\bm{\omega},\bm{v}^{j}\rangle\langle\bm{\omega},\delta\bm{v}^{j(2)}\rangle}{(\lambda_j)^2}-\frac{4\delta\lambda_j^{(1)}\langle\bm{\omega},\bm{v}^{j}\rangle\langle\bm{\omega},\delta\bm{v}^{j(1)}\rangle}{(\lambda_j)^3}+\frac{(3(\delta\lambda_j^{(1)})^2-2\lambda_j\delta\lambda_j^{(2)})\langle\bm{\omega},\bm{v}^{j}\rangle^2}{(\lambda_j)^4}\right)+\mathcal{O}(\epsilon^{3+\alpha_j}),\label{eq:05}
\end{align}
where $\alpha_j=-2$ when $j=2$, but is otherwise zero. Due to the the different scaling with $\epsilon$, the terms associated with $j=2$ are larger than those for $j\ge3$. Inserting Eq.~(\ref{eq:05}) into Eq.~(\ref{eq:04}) yields
\begin{align}
J(\bm{\omega},&L(\epsilon))=\epsilon^{-2}N^{-1}\left(\frac{\langle\bm{\omega},\bm{v}^2\rangle^2}{(\delta\lambda_2^{(1)})^2}\right)+\epsilon^{-1}N^{-1}\left(\frac{2\langle\bm{\omega},\bm{v}^2\rangle\langle\bm{\omega},\delta\bm{v}^{2(1)}\rangle}{(\delta\lambda_2^{(1)})^2}-\frac{2\delta\lambda_2^{(2)}\langle\bm{\omega},\bm{v}^2\rangle^2}{(\delta\lambda_2^{(1)})^3}\right)\nonumber\\
&+N^{-1}\left(\frac{\langle\bm{\omega},\delta\bm{v}^{2(1)}\rangle^2+2\langle\bm{\omega},\bm{v}^{2}\rangle\langle\bm{\omega},\delta\bm{v}^{2(2)}\rangle}{(\delta\lambda_2^{(1)})^2}-\frac{4\delta\lambda_2^{(2)}\langle\bm{\omega},\bm{v}^{2}\rangle\langle\bm{\omega},\delta\bm{v}^{2(1)}\rangle}{(\delta\lambda_2^{(1)})^3}+\frac{(3(\delta\lambda_2^{(2)})^2-2\delta\lambda_2^{(1)}\delta\lambda_2^{(3)})\langle\bm{\omega},\bm{v}^{2}\rangle^2}{(\delta\lambda_2^{(1)})^4}\right)\nonumber\\
&+\eta_1J(\bm{\omega}^1,L_1)+\eta_2J(\bm{\omega}^2,L_2)+\epsilon N^{-1}\sum_{j=3}^N\left(\frac{2\langle\bm{\omega},\bm{v}^j\rangle\langle\bm{\omega},\delta\bm{v}^{j(1)}\rangle}{(\lambda_j)^2}-\frac{2\delta\lambda_j^{(1)}\langle\bm{\omega},\bm{v}^j\rangle^2}{(\lambda_j)^3}\right)+\mathcal{O}(\epsilon N^{-1},\epsilon^2),\label{eq:06}
\end{align}
\end{widetext}
where we have used that 
$\frac{1}{N}\sum_{j=3}^N\frac{\langle\bm{\omega},\bm{v}^j\rangle^2}{\lambda_j^2}=\eta_1J(\bm{\omega}^{(1)},L^{(1)})+\eta_2J(\bm{\omega}^{(2)},L^{(2)})$ and $\eta_s=N_s/N$ is the fraction of nodes in subsystem $s\in\{1,2\}$. We note that Eq.~(\ref{eq:06}) diverges in the limit $\epsilon\to0$, as does Eq.~(\ref{eq:04}) and, in fact, so does the original SAF in Eq.~(\ref{eq:02}). However, in this limit the network becomes disconnected, so we are interested in the behavior of Eq.~(\ref{eq:06}) for finite, but small $\epsilon$. 

While Eq.~(\ref{eq:06}) may appear daunting, the key insight is that the inner product $\langle\bm{\omega},\bm{v}^2\rangle$ appears in several leading-order terms. Recalling the structure of $\bm{v}^2$, and writing $\bm{\omega}=[\bm{\omega}^{(1)},\bm{\omega}^{(2)}]^T$, where $\bm{\omega}^{(1)}$ and $\bm{\omega}^{(2)}$ are the frequency vectors corresponding to the two subsystems, we have that $\langle\bm{\omega},\bm{v}^2\rangle=\sqrt{\eta_1\eta_2}(\langle\bm{\omega}^{(1)}\rangle-\langle\bm{\omega}^{(2)}\rangle)$. Thus, if the subsystems' mean frequencies can be engineered to match, $\langle\bm{\omega}^{(1)}\rangle=\langle\bm{\omega}^{(2)}\rangle$, then many terms vanish to yield
\begin{align}
J&(\bm{\omega},L(\epsilon))=\eta_1J(\bm{\omega}^{(1)},L_1)+\eta_2J(\bm{\omega}^{(2)},L_2)\nonumber\\&+\epsilon N^{-1}\sum_{j=3}^N\left(\frac{2\langle\bm{\omega},\bm{v}^j\rangle\langle\bm{\omega},\delta\bm{v}^{j(1)}\rangle}{(\lambda_j)^2}-\frac{2\delta\lambda_j^{(1)}\langle\bm{\omega},\bm{v}^j\rangle^2}{(\lambda_j)^3}\right)\nonumber\\&+N^{-1}\frac{\langle\bm{\omega},\delta\bm{v}^{2(1)}\rangle^2}{(\delta\lambda_2^{(1)})^2}+\mathcal{O}(N^{-1}\epsilon,\epsilon^2),
\label{eq:07}
\end{align}
which has the leading order approximation
\begin{align}J(\bm{\omega},L(\epsilon))\approx\eta_1J(\bm{\omega}^{(1)},L_1)+\eta_2J(\bm{\omega}^{(2)},L_2) .
\end{align}
Thus, when the subsystems' mean frequencies are equal, or nearly equal, we find that the SAF of the full system can be approximated as the weighted average of the SAFs for the subsystems. A generalization of this theory is presented in Appendices A and B, and we discuss and utilize these results in the next section.

\begin{figure*}[t]
\centering
\epsfig{file =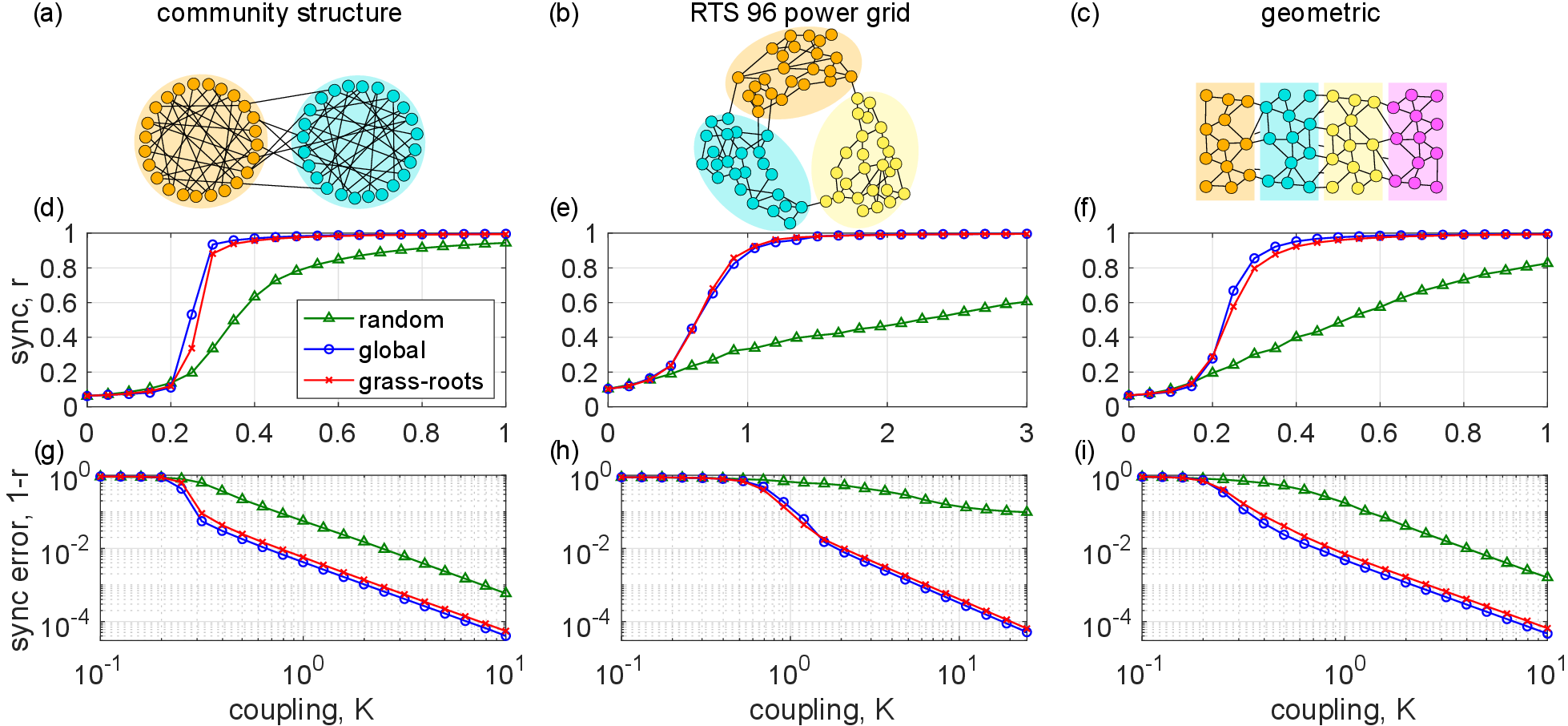, clip =,width=0.90\linewidth }
\caption{{\it Grass-roots optimization.} Illustrations of (a) a random network with two communities, (b) the IEEE RTS 96 power grid, and (c) a random geometric network. (d)--(f) The degree of synchronization $r$ and (g)--(i) synchronization error $1-r$ as a function of coupling strength $K$ for the three respective network types with either randomly allocated frequencies (green triangles), globally-optimized frequencies (blue circles), or grass-roots optimized frequencies (red crosses).}\label{fig2}
\end{figure*}

\subsection{Grass Roots Optimization of Phase Synchronization}\label{sec:2c}

We now present a method for grass-roots optimization of synchronization, including a multiscale mechanism in which subsystems coordinate local optimizations to optimize a system's global synchronization properties. Consider a network that can be partitioned into $C$ subsystems such that the adjacency matrix $A$ may be rewritten in a block form $A=A_D+B$, where $A_D=\text{diag}(A^{(1)},\dots,A^{(C)})$ is a block-diagonal matrix containing the subsystems' adjacency matrices, and the off-diagonal blocks of $B$ encode edges between subsystems. We assume that the blocks in $B$ are sparser than the diagonal blocks in $A_D$ and that the diagonal blocks in $B$ are matrices of zeros. For each subsystem $s$, we define its associated combinatorial Laplacian matrix $L^{(s)}$ and its associated vector $\bm{\omega}^{(s)}$ of frequencies. 

As we  show in the Appendices, {\it under the condition where the subsystems' mean oscillator frequencies are equal}, the SAF for the full system may be approximated by a linear combination of the subsystem-specific SAFs,
\begin{align}
J(\bm{\omega},L)\approx\eta_1J(\bm{\omega}^{(1)},L^{(1)})+\cdots+\eta_CJ(\bm{\omega}^{(C)},L^{(C)}),\label{eq:03}
\end{align}
where $\eta_s$ is the relative size of subsystem $s$. This result leads to the following multiscale mechanism for grass-roots optimization: 
\begin{itemize}
\item[(i)] \emph{Global balancing of subsystems}: achieve a balanced set of local mean frequencies across all $C$ subsystems, i.e., minimize $\text{max}_{s,s'}|\langle\bm{\omega}^{(s)}\rangle-\langle\bm{\omega}^{(s')}\rangle|$; 
\item[(ii)] \emph{Local optimization of subsystems}: optimize the local SAFs, i.e., minimize each $J(\bm{\omega}^{(s)},L^{(s)})$. 
\end{itemize}

These two steps are illustrated in Fig.~\ref{fig1}(c), where the network is divided into disjoint subsystems which are then balanced and separately optimized. This framework is flexible and fits a wide range of application-specific constraints. These two intuitive steps help fill the theoretical gap between existing global optimization theory and local heuristics that promote synchrony.

\section{Numerical Experiments}\label{sec:03}

In this section we present numerical experiments to highlight the utility of grass roots optimization for network synchronization. In Sec.~\ref{sec:03a} we show that globally optimized and and grass-roots optimized systems have similar synchronization properties. In Sec.~\ref{sec:03b} we show that grass-roots optimized networks have the added advantage of being robust to subsystem islanding or fragmentation. In Sec.~\ref{sec:03c} we highlight how the framework is also effective for optimizing a cardiac dynamics model that does not fit the precise form of Eq.~\eqref{eq:01}.

\subsection{Grass-Roots Optimization for Three Network Examples}\label{sec:03a}

We now illustrate the effectiveness of grass-roots optimization across three classes of networks: (i) networks with community structure (generated by the stochastic block model~\cite{Holland1983} with two communities of sizes $N^{(1;2)} = 100$ and mean intra- and inter-degrees $\langle k^{(1;2)}\rangle = 5$ and $\langle k^{(12)}\rangle = 1$); (ii) the RTS 96 power grid~\cite{Grigg1999IEEE}; (iii) and noisy geometric networks~\cite{Taylor2015NatComms} (with $N=200$ nodes placed randomly in a $4\times1$ box with $95\%$ of links placed between the closest possible nodes pairs and the other $5\%$ of links placed randomly, with a mean degree of $\langle k\rangle=8$). As shown in Figs.~\ref{fig2}(a)--(c), we partition the three classes of networks into two, three, and four subsystems, respectively. (The four subsystems of the geometric networks are defined by the $\pm$ sign combinations in the first two non-trivial eigenvectors of $L$.) For each network, we assume that natural frequencies are given and cannot be modified, but may be rearranged. Thus, a global balance between subsystems [step (i)] may be obtained by shuffling frequencies between subsystems, while the subsystems may be locally optimized [step (ii)] by then shuffling frequencies within each subsystem. To optimize each network, we use an accept-reject algorithm, proposing $5\times10^4$ switches between randomly chosen pairs of frequencies and accepting switches that decrease the SAF. 

In Figs.~\ref{fig2}(d)--(f), we plot $r$ vs $K$ for systems with randomly allocated (green triangles), globally optimized (blue circles), and grass-roots optimized (red crosses) frequencies for the three classes of networks. All data points are averaged across $50$ random networks and natural frequency realizations (drawn from the standard normal distribution) except for the power grid, where the same network is used. Note the comparably strong synchronization properties for both the global and grass-roots optimized cases, and that  sometimes grass-roots-optimized systems even exhibit stronger synchrony than the globally optimized systems due to the optimization algorithms' stochasticity. To differentiate the two cases we plot the synchronization error $1-r$ vs $K$ in a log-log scale in Figs.~\ref{fig2}(g)--(i), revealing that grass-roots optimization is effective across a wide range of network structures.

\begin{figure}[b]
\centering
\epsfig{file =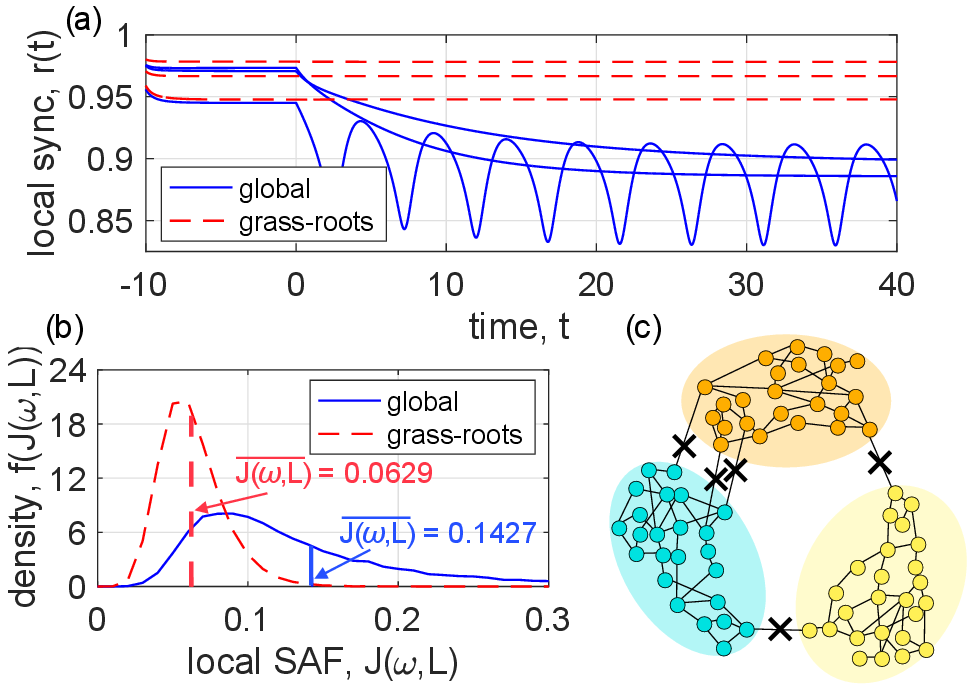, clip =,width=1.0\linewidth }
\caption{{\it Robustness to islanding and target attacks.} (a) Example of local (subsystem) order parameters for the RTS 96 power grid before and after islanding at $t=0$ for global (solid blue) and grass-roots (dashed red) optimization. (b) Density of local (subsystem) SAFs after islanding for global (solid blue) and grass-roots (dashed red) optimization. (c) Illustration of the islanded subsystems.}\label{fig4}
\end{figure}

\subsection{Application {to} Islanding of Power Grids}\label{sec:03b}

Here we highlight an advantage of grass-roots optimized networks versus globally optimized networks: they yield  networks whose synchronzation  properties are more robust to when subsystems are islanded or dismantled from one another. For instance, modern power grids feature microgrids--smaller subsystems that island (i.e., separate) themselves from the larger grid~\cite{Porco2013}. We predict such a feature to be advantageous in biological processes, which is a main motivator for our work. 

As an example, we consider the RTS 96 power grid before and after the islanding of three subsystems [illustrated in Fig.~\ref{fig4}(c)]. In Fig.~\ref{fig4}(a) we plot time series of the three local order parameters using global (solid blue) and grass-root (dashed red) optimization with $K=1$ and normally-distributed frequencies. Edges between subsystems are removed at time $t=0$. Before islanding ($t<0$) both cases display strong synchronization properties. After islanding ($t\ge0$) the globally-optimized system displays significantly weaker synchronization properties and a desynchronization event (indicated by oscillations). On the other hand, the grass-roots optimized system maintains its strong synchronization properties. This is further demonstrated in Fig.~\ref{fig4}(b), where we plot the density of local, i.e., subsystem-specific, SAFs for globally (solid blue) and grass-roots (dashed red) optimized systems obtained from $10^4$ realizations. We indicate the respective means $\overline{J(\bm{\omega}^{(s)},L^{(s)})}=0.1427$ and $0.0629$ of the local SAFs with vertical lines.

\subsection{Application {to} Cardiac Pacemakers}\label{sec:03c}

\begin{figure}[t]
\centering
\epsfig{file =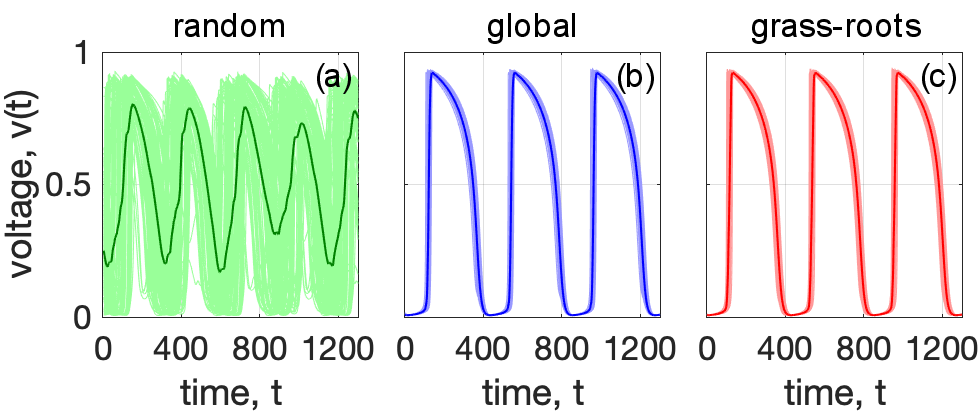, clip =,width=1.0\linewidth }
\caption{{\it Grass-roots optimization of cardiac pacemakers.} For a geometric network of $N=100$ cardiac pacemakers with two subsystems, the time series of non-dimensional voltage for (a) random, (b) globally optimized, and (c) grass-roots optimized allocations.}\label{fig3}
\end{figure}

Next we demonstrate that grass-roots optimization may be effectively used to optimize oscillator systems that do not fit the specific form of Eq.~\eqref{eq:01}. We study a model of cardiac pacemaker cells~\cite{Djabella2007IEEE} whose states $(v_i,h_i)$ for $i=1,\dots,N$ correspond to non-dimensional voltage and a gating variables that summarize ionic concentrations and evolve via
\begin{align}
\dot{v}_i&=\tau_i^{-1}f(v_i,h_i)+K_v\sum_{j=1}A_{ij}(v_j-v_i),\\
\dot{h}_i&=\tau_i^{-1}g(v_i,h_i)+K_h\sum_{j=1}A_{ij}(h_j-h_i),
\end{align}
where $f(v,h)=h(v+0.2)^2(1-v)/0.3-v/6$ and $g(v,h)=1/150+(8.333\times10^{-4})[1-\text{sgn}(v-0.13)]\{0.5[1-\text{sgn}(v-0.13)]-h\}$. The timescales $\tau_i$ represent local heterogeneity, scaling the period of each isolated cell, resulting in an effective natural frequency proportional to $\tau_i^{-1}$. We consider a geometric network of $N=100$ pacemakers with two subsystems, take $\tau_i^{-1}$ to be uniformly distributed in $[0.4,1.6]$, and use $K_v=0.0072$ and $K_h=0.0035$ (to indicate a stronger coupling via the voltage diffusion compared to ionic diffusion). We then implemented random, globally optimized, and grass-roots optimized allocations, plotting the resulting time series of voltage in Figs.~\ref{fig3}(a)--(c), respectively. Individual time series $v_i(t)$ are plotted lightly, while the mean is plotted with a dark stroke. Despite the stiff, nonlinear dynamics, both grass-roots and global optimization work remarkably well, yielding a strong, robust series of mean action potentials, while the random allocation does not.

\section{Discussion}\label{sec:07}

While recent progress has been made in optimizing collective behavior in complex systems, the resulting techniques and methodologies rely largely on global network information~\cite{Pecora1998PRL,Nishikawa2006PRE,Skardal2014PRL,Skardal2016Chaos,Taylor2016SIAM}. Given direct evidence of paracrine signaling, i.e., local communication, in biological systems and the likelihood that global information is unavailable, the collective, self-organizing processes by which naturally-occurring systems self-optimize remain an open critical question. Grass-roots optimization is a multiscale mechanism for coordinating and optimizing the local synchronization properties of a network's subsystems that provides a plausible mechanism for self-optimization in biological and other systems, such as cardiac pacemakers~\cite{Bychkov2020JACC} and genetic oscillators~\cite{Rajapakse2009PNAS}. It can also support the design of decentralized, parallelizable and scalable algorithms to engineer man-made systems that are robust to network dismantling. Notably, these very same features may have provided an evolutionary advantage for biological systems that crucially depend on synchronization. 


\acknowledgements
PRC acknowledges the Interdisciplinary Science and Summer Research Programs at Trinity College. DT acknowledges NSF grant DMS-2052720 and Simons Foundation award \#578333. PSS acknowledges NSF grant MCB-2126177.

\begin{widetext}

\appendix

\section{Local Approximation of the SAF for Networks with Three Subsystems}\label{app:01}

To provide insight into systems with more than two subsystems, we present here the case of three subsystems and derive a local approximation to the SAF analogous to the one which we presented in the main text. In this case the network adjacency matrix can be written in block form as
\begin{align}
A=\begin{bmatrix}A^{(1)} & B^{(12)} & B^{(13)} \\ B^{(12)T} & A^{(2)} & B^{(23)} \\ B^{(13)T} & B^{(23)T} & A^{(3)} \end{bmatrix},\label{eq:a01}
\end{align}
where $A^{(1)}$, $A^{(2)}$, and $A^{(3)}$ are the adjacency matrices for the three subsystems and $B^{(12)}$, $B^{(13)}$, and $B^{(23)}$ captures the connections between the respective subsystems. We denote the sizes of the three subsystems by $N_1$, $N_2$, and $N_3$ so that $A^{(1)}\in\mathbb{R}^{N_1\times N_1}$, $A^{(2)}\in\mathbb{R}^{N_2\times N_2}$, $A^{(3)}\in\mathbb{R}^{N_3\times N_3}$, $B^{(12)}\in\mathbb{R}^{N_1\times N_2}$, $B^{(13)}\in\mathbb{R}^{N_1\times N_3}$, and $B^{(23)}\in\mathbb{R}^{N_2\times N_3}$. We are interested then in the perturbed combinatorial Laplacian, given by
\begin{align}
L(\epsilon) = L_0+\epsilon\Delta L,\label{eq:a02}
\end{align}
where
\begin{align}
L_0=\begin{bmatrix}L^{(1)} & 0 & 0 \\ 0 & L^{(2)} & 0\\ 0 & 0 & L^{(3)}\end{bmatrix},\label{eq:a03}
\end{align}
$\Delta L = (\|L_0\|/\|L_B\|)L_B$, and
\begin{align}
L_B=\begin{bmatrix}D_{B^{(12)}+B^{(13)}} & -B^{(12)} & -B^{(13)} \\ -B^{(12)T} & D_{B^{(12)T}+B^{(23)}} & -B^{(23)}\\ -B^{(13)T} & -B^{(23)T} & D_{B^{(13)T}+B^{(23)T}}\end{bmatrix}.\label{eq:a04}
\end{align}
Once again, the choice $\epsilon=\|L_B\|/\|L_0\|\ll1$ recovers the original Laplacian matrix.

As in the two-subsystem case, it is useful to first discuss the spectral properties of $L_0$. Since it is a block-diagonal matrix, its eigenvalues are given by the union of the eigenvalues of the respective blocks,
\begin{align}
\{\lambda_j\}_{j=1}^N=\{\mu_j\}_{j=1}^{N_1}\bigcup\{\nu_j\}_{j=1}^{N_2}\bigcup\{\eta_j\}_{j=1}^{N_3},\label{eq:a05}
\end{align}
where $\{\mu_j\}_{j=1}^{N_1}$ denotes the eigenvalues of $L^{(1)}$, $\{\nu_j\}_{j=1}^{N_2}$ denotes the eigenvalues of $L^{(2)}$, and $\{\eta_j\}_{j=1}^{N_3}$ denotes the eigenvalues of $L^{(3)}$. The associated eigenvectors are given by
\begin{align}
\{\bm{v}^j\}_{j=1}^N=\left\{\begin{bmatrix}\bm{u}^j\\\bm{0}\\\bm{0}\end{bmatrix}\right\}_{j=1}^{N_1}\bigcup\left\{\begin{bmatrix}\bm{0}\\\bm{x}^j\\\bm{0}\end{bmatrix}\right\}_{j=1}^{N_2}\bigcup\left\{\begin{bmatrix}\bm{0}\\\bm{0}\\\bm{y}^j\end{bmatrix}\right\}_{j=1}^{N_3}.\label{eq:a06}
\end{align}
where $\{\bm{u}^j\}_{j=1}^{N_1}$, $\{\bm{x}^j\}_{j=1}^{N_2}$, and $\{\bm{y}^j\}_{j=1}^{N_3}$ are the associated eigenvectors for $L^{(1)}$, $L^{(2)}$, and $L^{(3)}$, respectively. The most critical observation to make is that each diagonal block of $L_0$ has a trivial eigenvalue, namely, $\mu_1,\nu_1,\eta_1=0$, so the nullspace of $L_0$ is three-dimensional since it has a triple eigenvalue degeneracy at $\lambda_{1,2,3}=0$. It is then convenient to rewrite the basis vectors for this trivial eigenspace using the following eigenvectors:
\begin{align}
\bm{v}^1=\frac{1}{\sqrt{N}}\begin{bmatrix}\bm{1}\\\bm{1}\\\bm{1}\end{bmatrix},~~~\bm{v}^2=\frac{\sqrt{N_1N_2}}{N_1+N_2}\begin{bmatrix}\bm{1}/N_1\\-\bm{1}/N_2\\\bm{0}\end{bmatrix},~~~\bm{v}^3=\frac{\sqrt{N_2N_3}}{N_2+N_3}\begin{bmatrix}\bm{0}\\\bm{1}/N_2\\-\bm{1}/N_3\end{bmatrix},\label{eq:a07}
\end{align}
where, similar to the two subsystem case, $\bm{v}^1$ is the constant-valued eigenvector that is associated with the synchronization manifold and whose eigenvalue $\lambda_1=0$ remains constant as $\epsilon$ increases (i.e., $v^1(\epsilon) = v^1$ regardless of $\epsilon$). On the other hand, $\bm{v}^2$ and $\bm{v}^3$ will play important roles in the perturbation analysis since $\lambda_2(\epsilon)$ and $\lambda_3(\epsilon)$ must take positive values for any $\epsilon>0$. We note that the vector $\sqrt{N_1N_3}/(N_1+N_3)\begin{bmatrix}\bm{1}/N_1\\ \bm{0} \\ -\bm{1}^T/N_3\end{bmatrix}$ may also be used in place of either $\bm{v}^2$ or $\bm{v}^3$, but as it is just a linear combination of the two vectors already chosen, it yields the same results given below. 

Given the initial spectral properties of $L_0$, we consider the following perturbative expansions. Specifically, for the eigenvalues of $L(\epsilon)$ we have
\begin{align}
\lambda_j(\epsilon)&=\epsilon\delta\lambda_j^{(1)}+\epsilon^2\delta\lambda_j^{(2)}+\mathcal{O}(\epsilon^3),\label{eq:a08}
\end{align}
for $j=2,3$ and 
\begin{align}
\lambda_j(\epsilon)&=\lambda_j+\epsilon\delta\lambda_j^{(1)}+\epsilon^2\delta\lambda_j^{(2)}+\mathcal{O}(\epsilon^3),\label{eq:a09}
\end{align}
for $j=4,\dots,N$. We again assume that the eigenvectors of $L(\epsilon)$ are continuously differentiable to approximate
\begin{align}
\bm{v}^j(\epsilon)&=\bm{v}^j+\epsilon\delta\bm{v}^{j(1)}+\epsilon^2\delta\bm{v}^{j(2)}+\mathcal{O}(\epsilon^3).\label{eq:a10}
\end{align}
for $j=2,\dots,N$.

Our primary interest is the SAF of the perturbed network, and as we did in the two subsystem case with the term associated with $j=2$, here we will treat the terms associated with $j=2$ and $3$ separately:
\begin{align}
J(\bm{\omega},L(\epsilon))=\frac{1}{N}\left(\frac{\langle\bm{\omega},\bm{v}^2(\epsilon)\rangle}{\lambda_2(\epsilon)}\right)^2+\frac{1}{N}\left(\frac{\langle\bm{\omega},\bm{v}^3(\epsilon)\rangle}{\lambda_3(\epsilon)}\right)^2+\frac{1}{N}\sum_{j=4}^N\left(\frac{\langle\bm{\omega},\bm{v}^j(\epsilon)\rangle}{\lambda_j(\epsilon)}\right)^2.\label{eq:a11}
\end{align}
We now consider the contribution of these different terms. Beginning with the terms associated with $j=2$ and $3$, insert Eqs.~(\ref{eq:a08}) and (\ref{eq:a10}) into the relevant terms in Eq.~(\ref{eq:a11}), expand, and collect similar terms to obtain
\begin{align}
\frac{1}{N}\left(\frac{\langle\bm{\omega},\bm{v}^j(\epsilon)\rangle}{\lambda_j(\epsilon)}\right)^2&=N^{-1}\epsilon^{-2}\left(\frac{\langle\bm{\omega},\bm{v}^j\rangle^2}{(\delta\lambda_j^{(1)})^2}\right)+N^{-1}\epsilon^{-1}\left(\frac{2\langle\bm{\omega},\bm{v}^j\rangle\langle\bm{\omega},\delta\bm{v}^{j(1)}\rangle}{(\delta\lambda_j^{(1)})^2}-\frac{2\delta\lambda_j^{(2)}\langle\bm{\omega},\bm{v}^j\rangle^2}{(\delta\lambda_j^{(1)})^3}\right)\nonumber\\
&+N^{-1}\left(\frac{\langle\bm{\omega},\delta\bm{v}^{j(1)}\rangle^2+2\langle\bm{\omega},\bm{v}^{j}\rangle\langle\bm{\omega},\delta\bm{v}^{j(2)}\rangle}{(\delta\lambda_j^{(1)})^2}-\frac{4\delta\lambda_j^{(2)}\langle\bm{\omega},\bm{v}^{j}\rangle\langle\bm{\omega},\delta\bm{v}^{j(1)}\rangle}{(\delta\lambda_j^{(1)})^3}\right.\nonumber\\
&~~~~~~~~~~~~~~~\left.+\frac{(3(\delta\lambda_j^{(2)})^2-2\delta\lambda_j^{(1)}\delta\lambda_j^{(3)})\langle\bm{\omega},\bm{v}^{j}\rangle^2}{(\delta\lambda_j^{(1)})^4}\right)+\mathcal{O}(N^{-1}\epsilon).\label{eq:a12}
\end{align}
On the other hand, for $j=4,\dots,N$, we insert Eqs.~(\ref{eq:a09}) and (\ref{eq:a10}) into the relevant terms in Eq.~(\ref{eq:a11}), expand, and collect similar terms to obtain
\begin{align}
\frac{1}{N}\left(\frac{\langle\bm{\omega},\bm{v}^j(\epsilon)\rangle}{\lambda_j(\epsilon)}\right)^2&=N^{-1}\left(\frac{\langle\bm{\omega},\bm{v}^j\rangle^2}{(\lambda_j)^2}\right)+N^{-1}\epsilon\left(\frac{2\langle\bm{\omega},\bm{v}^j\rangle\langle\bm{\omega},\delta\bm{v}^{j(1)}\rangle}{(\lambda_j)^2}-\frac{2\delta\lambda_j^{(1)}\langle\bm{\omega},\bm{v}^j\rangle^2}{(\lambda_j)^3}\right)\nonumber\\
&+N^{-1}\epsilon^2\left(\frac{\langle\bm{\omega},\delta\bm{v}^{j(1)}\rangle^2+2\langle\bm{\omega},\bm{v}^{j}\rangle\langle\bm{\omega},\delta\bm{v}^{j(2)}\rangle}{(\lambda_j)^2}-\frac{4\delta\lambda_j^{(1)}\langle\bm{\omega},\bm{v}^{j}\rangle\langle\bm{\omega},\delta\bm{v}^{j(1)}\rangle}{(\lambda_j)^3}\right.\nonumber\\
&~~~~~~~~~~~~~~~\left.+\frac{(3(\delta\lambda_j^{(1)})^2-2\lambda_j\delta\lambda_j^{(2)})\langle\bm{\omega},\bm{v}^{j}\rangle^2}{(\lambda_j)^4}\right)+\mathcal{O}(N^{-1}\epsilon^3).\label{eq:a13}
\end{align}
Inserting Eqs.~(\ref{eq:a12}) and (\ref{eq:a13}) into Eq.~(\ref{eq:a11}), we then obtain
\begin{align}
J(&\bm{\omega},L(\epsilon))=N^{-1}\epsilon^{-2}\left(\frac{\langle\bm{\omega},\bm{v}^2\rangle^2}{(\delta\lambda_2^{(1)})^2}+\frac{\langle\bm{\omega},\bm{v}^3\rangle^2}{(\delta\lambda_3^{(1)})^2}\right)\nonumber\\
&+N^{-1}\epsilon^{-1}\left(\frac{2\langle\bm{\omega},\bm{v}^2\rangle\langle\bm{\omega},\delta\bm{v}^{2(1)}\rangle}{(\delta\lambda_2^{(1)})^2}-\frac{2\delta\lambda_2^{(2)}\langle\bm{\omega},\bm{v}^2\rangle^2}{(\delta\lambda_2^{(1)})^3}+\frac{2\langle\bm{\omega},\bm{v}^3\rangle\langle\bm{\omega},\delta\bm{v}^{3(1)}\rangle}{(\delta\lambda_3^{(1)})^2}-\frac{2\delta\lambda_3^{(2)}\langle\bm{\omega},\bm{v}^3\rangle^2}{(\delta\lambda_3^{(1)})^3}\right)\nonumber\\
&+N^{-1}\left(\frac{\langle\bm{\omega},\delta\bm{v}^{2(1)}\rangle^2+2\langle\bm{\omega},\bm{v}^{2}\rangle\langle\bm{\omega},\delta\bm{v}^{2(2)}\rangle}{(\delta\lambda_2^{(1)})^2}-\frac{4\delta\lambda_2^{(2)}\langle\bm{\omega},\bm{v}^{2}\rangle\langle\bm{\omega},\delta\bm{v}^{2(1)}\rangle}{(\delta\lambda_2^{(1)})^3}+\frac{(3(\delta\lambda_2^{(2)})^2-2\delta\lambda_2^{(1)}\delta\lambda_2^{(3)})\langle\bm{\omega},\bm{v}^{2}\rangle^2}{(\delta\lambda_2^{(1)})^4}\right.\nonumber\\
&~~~~~~~~~~~~+\left.\frac{\langle\bm{\omega},\delta\bm{v}^{3(1)}\rangle^2+2\langle\bm{\omega},\bm{v}^{3}\rangle\langle\bm{\omega},\delta\bm{v}^{3(2)}\rangle}{(\delta\lambda_3^{(1)})^2}-\frac{4\delta\lambda_3^{(2)}\langle\bm{\omega},\bm{v}^{3}\rangle\langle\bm{\omega},\delta\bm{v}^{3(1)}\rangle}{(\delta\lambda_3^{(1)})^3}+\frac{(3(\delta\lambda_3^{(2)})^2-2\delta\lambda_3^{(1)}\delta\lambda_3^{(3)})\langle\bm{\omega},\bm{v}^{3}\rangle^2}{(\delta\lambda_3^{(1)})^4}\right)\nonumber\\
&+\eta_1J(\bm{\omega}^1,L_1)+\eta_2J(\bm{\omega}^2,L_2)+\eta_3J(\bm{\omega}^2,L_3)+\epsilon\left[N^{-1}\sum_{j=4}^N\left(\frac{2\langle\bm{\omega},\bm{v}^j\rangle\langle\bm{\omega},\delta\bm{v}^{j(1)}\rangle}{(\lambda_j)^2}-\frac{2\delta\lambda_j^{(1)}\langle\bm{\omega},\bm{v}^j\rangle^2}{(\lambda_j)^3}\right)\right]+\mathcal{O}(N^{-1}\epsilon,\epsilon^2),\label{eq:a14}
\end{align}
where we have used that, for the three subsystem case, we have
\begin{align}
\frac{1}{N}\sum_{j=4}^N\frac{\langle\bm{\omega},\bm{v}^j\rangle^2}{\lambda_j^2}=\eta_1J(\bm{\omega}^1,L_1)+\eta_2J(\bm{\omega}^2,L_2)+\eta_3J(\bm{\omega}^3,L_3).\label{eq:a15}
\end{align}

Lastly, to complete the analysis we consider not only the contributions of $\langle\bm{\omega},\bm{v}^2\rangle$, but also $\langle\bm{\omega},\bm{v}^3\rangle$. In particular, we note that
\begin{align}
\langle\bm{\omega},\bm{v}^2\rangle=\frac{\sqrt{\eta_1\eta_2}}{\eta_{12}}(\langle\omega^1\rangle-\langle\omega^2\rangle),\label{eq:a16}
\end{align}
and
\begin{align}
\langle\bm{\omega},\bm{v}^3\rangle=\frac{\sqrt{\eta_2\eta_3}}{\eta_{23}}(\langle\omega^2\rangle-\langle\omega^3\rangle),\label{eq:a17}
\end{align}
where $\eta_{ij}=(N_i+N_j)/N$. Thus, if we may engineer the network such that $\langle\omega^1\rangle=\langle\omega^2\rangle=\langle\omega^3\rangle$, then all terms in Eq.~(\ref{eq:a14}) with $\langle\bm{\omega},\bm{v}^2\rangle$ or $\langle\bm{\omega},\bm{v}^3\rangle$ vanish, yielding
\begin{align}
J(\bm{\omega},L(\epsilon))&=\eta_1J(\bm{\omega}^1,L_1)+\eta_2J(\bm{\omega}^2,L_2)+\eta_3J(\bm{\omega}^2,L_3)+N^{-1}\left(\frac{\langle\bm{\omega},\delta\bm{v}^{2(1)}\rangle^2}{(\delta\lambda_2^{(1)})^2}+\frac{\langle\bm{\omega},\delta\bm{v}^{3(1)}\rangle^2}{(\delta\lambda_3^{(1)})^2}\right)\nonumber\\
&+\epsilon\left[N^{-1}\sum_{j=4}^N\left(\frac{2\langle\bm{\omega},\bm{v}^j\rangle\langle\bm{\omega},\delta\bm{v}^{j(1)}\rangle}{(\lambda_j)^2}-\frac{2\delta\lambda_j^{(1)}\langle\bm{\omega},\bm{v}^j\rangle^2}{(\lambda_j)^3}\right)\right]+\mathcal{O}(N^{-1}\epsilon,\epsilon^2),\label{eq:a18}
\end{align}
where the leading-order behavior of the perturbed SAF is simply given by a weighted average of the subsystem-specific SAFs and the weights come from their relative sizes, which is our desired result and the analogous version of Eq.~(7) in the main text.

\section{Local Approximation of the SAF for Networks with an Arbitrary Number of Subsystems}\label{app:02}

Before concluding, we emphasize that the three subsystem case above informs the generalization of the local approximation to an arbitrary number of subsystems. In particular, for $C$ subsystems, the unperturbed Laplacian $L_0$ will contain $C$ diagonal blocks, each with a trivial eigenvalue. Thus, a basis for the trivial eigenspace must be chosen so that, in addition to $\bm{v}^1\propto\bm{1}$, there are $C-1$ eigenvectors whose eigenvalues will becomes positive for positive $\epsilon$. This can be done by choosing, for instance,
\begin{align}
\bm{v}^2=\begin{bmatrix}\bm{1}/N_1\\-\bm{1}/N_2\\ \bm{0} \\ \vdots \\ \bm{0}\end{bmatrix},~~
\bm{v}^3=\begin{bmatrix}\bm{0} \\ \bm{1}/N_2\\-\bm{1}/N_3 \\ \vdots \\ \bm{0}\end{bmatrix},~~
\cdots ~~, ~~ 
\bm{v}^j=\begin{bmatrix}\vdots \\ \bm{1}/N_{j-1}\\-\bm{1}/N_j\\ \vdots \\ \bm{0}\end{bmatrix},~~
\cdots ~~,~~
\bm{v}^C=\begin{bmatrix} \bm{0} \\ \vdots \\ \bm{0} \\ \bm{1}/N_{C-1}\\-\bm{1}/N_C\end{bmatrix}.
\end{align}
Then, after expansion, setting $\langle\omega^1\rangle=\cdots=\langle\omega^C\rangle$ causes the two lowest order contributions to $J(\bm{\omega},L(\epsilon))$ originating from the terms associated with $j=2,\dots,C$ to vanish, yielding, to leading order,
\begin{align}
J(\bm{\omega},L(\epsilon))\approx\eta_1J(\bm{\omega}^1,L^{(1)})+\cdots+\eta_CJ(\bm{\omega}^C,L^{(C)}).\label{eq:a20}
\end{align}

\end{widetext}

\bibliographystyle{plain}

\end{document}